\documentclass[journal]{IEEEtran}
\usepackage{cite}
\usepackage{caption}
\usepackage{graphicx}
\usepackage{psfrag}
\DeclareGraphicsExtensions{.xps}
\usepackage{epstopdf} 
\usepackage{subcaption}
\usepackage[cmex10]{amsmath}
\usepackage{amssymb}
\usepackage{color}
\usepackage{amsthm}

\usepackage{tikz}
\usetikzlibrary{positioning}
\usetikzlibrary{shapes,arrows}
\tikzstyle{line}=[draw] 
\usepackage{pgfplots}
\usepackage{multirow}
\usepackage{algorithm}% http://ctan.org/pkg/algorithms
\usepackage{algpseudocode}% http://ctan.org/pkg/algorithmicx
\algrenewcommand{\algorithmiccomment}[1]{\hskip3em$\bullet$ #1} % this is to change the comment shape
\newlength\myindent
\makeatletter
\newcommand\fs@nobottomruled{\def\@fs@cfont{\bfseries}\let\@fs@capt\floatc@ruled
	\def\@fs@pre{\hrule height.8pt depth0pt \kern2pt}%
	\def\@fs@post{\kern5pt \hrule}% Formerly \def\@fs@post{\kern2pt\hrule\relax}%
	\def\@fs@mid{\kern2pt\hrule\kern2pt}%
	\let\@fs@iftopcapt\iftrue}
\makeatother

\floatstyle{nobottomruled}
\restylefloat{algorithm}

\begin{document}

\title{\huge Physical Adversarial Attacks Against End-to-End Autoencoder Communication Systems}

\author{Meysam Sadeghi and Erik G. Larsson,
	\thanks{ \newline \indent The authors (m.sadeghee@gmail.com, erik.g.larsson@liu.se) are with Department of Electrical Engineering (ISY), Link\"{o}ping University, Link\"{o}ping, Sweden. 
		
	The codes are available at \cite{code_mey_ae}.}  }

\maketitle

\normalsize
\begin{abstract}
	We show that end-to-end learning of communication systems through deep neural network (DNN) autoencoders can be extremely vulnerable to \emph{physical} adversarial attacks. Specifically, we elaborate how an attacker can craft effective  \emph{physical}  black-box adversarial attacks.  Due to the openness (broadcast nature) of the wireless channel, an adversary transmitter can increase the block-error-rate of a communication system by orders of magnitude by transmitting a well-designed perturbation signal over the channel. We reveal that the adversarial attacks are more destructive than jamming attacks. We also show that classical coding schemes are more robust than autoencoders against both adversarial and jamming attacks. The codes are available at \cite{code_mey_ae}.
\end{abstract}

\begin{IEEEkeywords}
	Adversarial attacks, Autoencoder systems, Deep learning, Wireless security, End-to-end learning, Security and Robustness of Deep Learning for Wireless Communications.
\end{IEEEkeywords}

\section{Introduction}

Deep neural networks (DNNs), due to their promising performance, are becoming an integral tool in many new disciplines \cite{goodfellow2016deep}. This has created a new series of applications and concepts in wireless communications under a new paradigm, namely deep learning based wireless communications. One such concept is the end-to-end learning of communication systems using autoencoders \cite{TimDL4Phy}. As the name suggests, this approach enables  end-to-end optimization and design of communication systems, which can potentially provide gains over the contemporary modularized design of these systems.

DNNs, despite their promising performance, are extremely susceptible to the so-called adversarial attacks \cite{goodfellow2014explaining,szegedy2013intriguing}. In adversarial attacks, the attacker adds a (small) perturbation $\mathbf{p}$ to the input of a DNN that causes erroneous outputs \cite{goodfellow2014explaining}. In contrast to the conventional jamming attacks, this perturbation is not noise but a deliberately optimized vector in the feature space of the input domain that can fool the model. This property of DNN has raised major concerns regarding their security and robustness.

Adversarial attacks can be classified based on the nature of  the attack. They can specifically be
divided into white-box and black-box attacks, based on the
amount of knowledge that the adversary has about the underlying NN \cite{Papernot2016Transferability}.
In white-box attacks, the adversary has the full knowledge of
the classifier, while in black-box attacks the adversary has no or limited  knowledge  of the
classifier \cite{Papernot2016Transferability}. Adversarial attacks can also be classified into digital and
physical attacks based on the degree of freedom that the
adversary has with respect to its access to the input of the
system \cite{kurakin2016adversarial}. In digital attacks, the adversary can \emph{precisely} design
the input of the model, while in physical attacks, the adversary indirectly applies the input to the model \cite{kurakin2016adversarial}. 

Recently digital adversarial attacks on modulation classification was studied in \cite{meysam_modcls}. Therein, digital black-box attacks were developed that can significantly reduce the performance of the DNN based classifier. However, given the  inherent random nature of the  wireless channel between the attacker and the receiver
in a wireless system, the feasibility of a digital attack might be questioned. But physical attacks are more practical.

In this letter, we study  \emph{physical} adversarial attacks against end-to-end autoencoder communication systems, and provide the following contributions. First, we present new algorithms for crafting \emph{effective physical} black-box adversarial attacks.\footnote{Algorithmically, the method for perturbation generation suggested here departs significantly from that in \cite{meysam_modcls}. There is a superficial similarity in that both use the SVD as a building block, but the problems and solutions are different.} Second, we reveal that the adversarial attacks are more destructive than jamming attacks. Third, we show  that classical communication schemes are more robust than the end-to-end autoencoder systems.

\section{System Model}
An autoencoder communication system has three main blocks, 1) transmitter, 2) channel, and 3) receiver \cite{TimDL4Phy}. The system receives an input message $s \in \mathcal{M}=\{1,2,\ldots,M \}$, where $M = 2^{k}$ is the dimension of $\mathcal{M}$ with $k$ being the number of bits per message. The message is then passed to the transmitter, where it applies a transformation $f:\mathcal{M} \to \mathbb{R}^{2n}$~\footnote{Note that the output of the transmitter is an $n$ dimensional complex vector which is transformed to a $2n$ real vector.} to the message $s$ to generate the transmitted signal $\mathbf{x} = f(s) \in \mathbb{R}^{2n}$. Similar to \cite{TimDL4Phy}, the transmitter enforces an average power constraint $\mathbb{E} [\vert x_i^2 \vert ] \leq 0.5 \; \forall i$.  Next, the signal $\mathbf{x}$ is sent to the receiver using the channel $n$ times. In this work, we consider an additive white Gaussian noise (AWGN) channel. Then, the receiver receives a signal $\mathbf{y}$, which contains the original signal $\mathbf{x}$, noise and other sources of interference. The receiver applies the transformation  $g:\mathbb{R}^{2n} \to \mathcal{M}$ to create $\hat{s} = g(\mathbf{y})$ which is the estimate of the transmitted message $s$. To enable the comparability of the results developed in different sections, we set $n=7$ and $k=4$. This also enables us to compare our results with the classical Hamming code (7,4).

\begin{figure}[h]
	\centering
	\includegraphics[width= 0.9 \columnwidth,trim={5.5cm 8.5cm 5.5cm 9.7cm} ,clip]{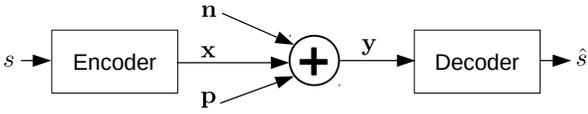}% trim={5.5cm 3.5cm 5.5cm 3cm}
	\caption{An adversarial attack against an autoencoder system.}
	\label{fig1}
\end{figure}

In deep learning terminology the transmitter and receiver are respectively called encoder and decoder and they are implemented using neural networks  \cite{goodfellow2016deep}. The input signal $s$ first passes through the encoder. Then the output of the encoder ($\mathbf{x}$) goes through the channel, where the attacker perturbation signal ($\mathbf{p}$) and the noise ($\mathbf{n}$) are added to the $\mathbf{x}$. Therefore the output of the channel will be $\mathbf{y} = \mathbf{x} + \mathbf{p} + \mathbf{n}$. Then $\mathbf{y}$ is fed to the decoder. This procedure is illustrated in Fig. \ref{fig1}.

To enable a benchmark for comparison, we use the same autoencoder structure as \cite{TimDL4Phy}, which is an MLP autoencoder.\footnote{The selected set of hyperparameters is the same as in \cite{TimDL4Phy}, to enable a direct comparison, and we do not claim their optimality. There are infinitely many combinations of hyperparameters that could be tested, but modifying  them is not  likely to change the conclusions, as   DNNs are known to be inherently   vulnerable to   adversarial attacks, see \cite{goodfellow2014explaining,kurakin2016adversarial,moosavi2017universal}.} Also, we use a convolutional neural network (CNN) based autoencoder to introduce the black-box adversarial attacks. The structure of these networks is given in Table~\ref{table1}.\footnote{We also investigated some variations of the network structure, e.g., ReLU activation, depth, different number of neurons. The results are quantitatively similar. All the codes are available at \cite{code_mey_ae}.} Detailed explanations of each element of Table~\ref{table1} can be found in \cite{goodfellow2016deep,TimDL4Phy}. We train the autoencoder in an end-to-end manner using the Adam optimizer, on the set of all possible messages $s \in \mathcal{M}$, using the cross-entropy loss function.

\begin{table}[]
	\begin{tabular}{l|lc|lc}
		& \multicolumn{2}{c|}{MLP Autoencoder}                                                                                                           & \multicolumn{2}{c}{CNN Autoencoder}                                                                                                         \\ \hline
		\multicolumn{1}{c|}{\begin{tabular}[c]{@{}c@{}}Block\\ Name\end{tabular}} & \begin{tabular}[c]{@{}l@{}}Layer\\ Name\end{tabular}              & \multicolumn{1}{l|}{\begin{tabular}[c]{@{}l@{}}Output\\ Dim.\end{tabular}} & \begin{tabular}[c]{@{}l@{}}Layer\\ Nmae\end{tabular}             & \multicolumn{1}{l}{\begin{tabular}[c]{@{}l@{}}Output\\ Dim.\end{tabular}} \\ \hline
		\multirow{5}{*}{Encoder}                                                  & Input                                                             & M                                                                          & Input                                                            & M                                                                         \\
		& Dense+eLU                                                         & M                                                                          & Dense + eLU                                                      & M                                                                         \\
		& Dense+Linear                                                      & 2n                                                   & conv1d+Flattening                                                & \multicolumn{1}{l}{16 $\times$ M}                                                \\
		& Normalization                                                     & 2n                                                                         & Dense + Linear                                                   & 2n                                                                        \\
		&                                                                   &                                                                            & Normalization                                                    & 2n                                                                        \\ \hline
		Channel                                                                   & \begin{tabular}[c]{@{}l@{}}noise \\ (+ perturbation)\end{tabular} & 2n                                                                         & \begin{tabular}[c]{@{}l@{}}noise\\ (+ perturbation)\end{tabular} & 2n                                                                        \\ \hline
		\multirow{4}{*}{Decoder}                                                  & Dense + ReLU                                                      & M                                                                          & conv2d                                                           & 16 $\times$ 2n                                                                     \\
		& Dens+Softmax                                                      & M                                                                          & conv2d+Flattening                                                & 8 $\times$ 2n                                                                      \\
		&                                                                   &                                                                            & Dense + ReLU                                                     & 2M                                                                         \\
		&                                                                   &                                                                            & Dens + Softmax                                                   & M                                                                        
	\end{tabular}
	\caption{The structure of considered autoencoders.}
	\label{table1}
\end{table}

\section{Crafting a White-box Adversarial Attack}

Given the structure of an autoencoder in Fig. \ref{fig1}, the adversary uses the broadcast nature of the channel and attacks the decoder block of the autoencoder communication system. To model the attack mathematically, let us denote the underlying DNN classifier at the decoder by $g:\mathbb{R}^{2n} \to \mathcal{M}$. Also denote $\mathbf{w} = \mathbf{x} + \mathbf{n}$. The classifier generates an output $\hat{s} = g(\mathbf{y})$ for every input $\mathbf{y} = \mathbf{w} + \mathbf{p}$. Note that when the attacker is absent $\mathbf{p}=\mathbf{0}$ and $\mathbf{w} = \mathbf{y}$. Given these definitions, an adversarial perturbation $\mathbf{p}$ for the considered autoencoder is \cite{goodfellow2014explaining, szegedy2013intriguing} 
\begin{align}
\label{one}
\min_{\mathbf{p}} & \; \; \Vert \mathbf{p} \Vert_{2}
\\
\text{s.t.} & \; \; g(\mathbf{w} + \mathbf{p}) \neq g(\mathbf{w})  \notag
\end{align}
where $\mathbf{n} \sim \mathcal{CN}(\mathbf{0}, \sigma^{2} \mathbf{I} )$ with $\sigma^{2} = N_{0} / 2RE_{b}$, $E_{b}$ is energy per bit, and $R = k/n$ is the rate in bits per channel use.

Solving \eqref{one} is challenging as $g$ does not
have a convex structure \cite{goodfellow2014explaining}. Therefore,
a common approach is to use the so-called fast gradient method (FGM)
\cite{szegedy2013intriguing,goodfellow2014explaining}, which finds
an approximate solution to \eqref{one}. The main idea is as
follows. For the input label $s \in \mathcal{M}$, denote the loss
function of the autoencoder by $L(\mathbf{w}+\mathbf{p},
s)$. Then, FGM uses a Taylor expansion of the loss function, i.e.,
$L(\mathbf{w}+\mathbf{p}, s) \approx L(\mathbf{w}, s) +
\mathbf{p}^{T} \nabla_{\mathbf{w}} L(\mathbf{w}, s)$, and sets
$\mathbf{p} = \alpha \nabla_{\mathbf{w}} L(\mathbf{w}, s)$, where
$\alpha$ is a scaling coefficient. Hence, FGM effectively
increases the loss function and provides an approximation of
$\mathbf{p}$, given that the input ($s$) is
known.\footnote{Further details  and more
	advanced methods can be found in
	\cite{szegedy2013intriguing,goodfellow2014explaining,meysam_modcls}.}

An adversarial attacker cannot directly apply the FGM
method to attack an autoencoder communication system, as the FGM
method requires knowledge of the input to the autoencoder, i.e.,
$s$, to create a perturbation $\mathbf{p}$. However, the attacker
does not know what symbol is being transmitted. To address this
issue, Alg. \ref{alg1} presents an iterative method to craft a
universal (i.e., input-agnostic) perturbation $\mathbf{p}$ that can
fool the autoencoder independently of its input. The main idea comes
from the literature on computer vision \cite{moosavi2017universal},
where it has been shown that by iteratively finding image-specific perturbations one can create image-agnostic perturbations.

\begin{algorithm}[t]
	\caption{ Crafting Physical Adversarial Perturbations \label{alg1}}
	\begin{algorithmic}[1]
		\Statex Inputs: 
		\begin{itemize}
			\item The full autoecoder's model, e.g., $g$ and the parameters of the decoder's underlying NN. %\vspace{-0.2em}
			\item The desired perturbation power $p_{power}$.  %\vspace{-0.1em}
			\item The variance of the channel noise $\sigma^{2}$.  %\vspace{-0.1em}
		\end{itemize} 
		\Statex Output: An input-agnostic adversarial perturbation, i.e., $\mathbf{p}$ %\vspace{-0.3em}
		%\vspace{-0.5em} 
		\noindent\rule{8.2cm}{0.2pt}
		
		\State Initialize: 
		%\vspace{-0.5em}
		\begin{itemize}
			\item Set \textit{number-of-samples} $ \leftarrow 10$. %\vspace{-0.2em}
			\item Set $\mathbf{p} \leftarrow \mathbf{0}$.  
		\end{itemize} 
		\For{\textit{i} in \textit{range}(\textit{number-of-samples})}
		\State Choose an $s \in \{1,2,\ldots,M\}$ uniformly at random.
		\State Create a random noise $\mathbf{n} \sim \mathcal{CN}(\mathbf{0},\sigma^2 \mathbf{I})$. 
		\State Set $\hat{s} \leftarrow g\left( f(s) + \mathbf{p} + \mathbf{n} \right)$.
		\If{$\hat{s} == s$}
		\State Set $\mathbf{w} \leftarrow \mathbf{w} + \mathbf{p}$ .
		\State Solve \eqref{one} and denote its solution by $\mathbf{p}_{update}$.
		\If{$\Vert \mathbf{p} + \mathbf{p}_{update} \Vert^{2}_{2}  \leq p_{power} $}
		\State $\mathbf{p} \leftarrow \mathbf{p} + \mathbf{p}_{update}$ 
		
		\Else
		\State $\mathbf{p} \leftarrow \sqrt{p_{power}} \; \dfrac{\mathbf{p} + \mathbf{p}_{update}}{\Vert \mathbf{p} + \mathbf{p}_{update} \Vert_{2}}$
		\EndIf 
		\EndIf
		\EndFor		
	\end{algorithmic}
	\vspace{-0.5em}
\end{algorithm}

In the first line of the Alg. \ref{alg1}, we
initialize the number of samples and the perturbation
$\mathbf{p}$. The number of samples determines the number of
iterations we use in order to create a universal perturbation. Then,
we repeatedly apply the following steps. First, we select an input
$s$ uniformly at random and pass it through the encoder. Next, we
compute the signal seen by the decoder, by adding the perturbation
and the noise to the output of encoder. Now we apply the decoder
block and obtain an estimate of the input, i.e., $\hat{s}$. If the
decoder correctly classifies the input message, we set $\mathbf{w} +
\mathbf{p}$ as $\mathbf{w}$. Using the new value of $\mathbf{w}$ and
by solving \eqref{one}, we search for an update perturbation $
\mathbf{p}_{update}$ that enforces misclassification of the new
input. Then we update the existing perturbation $\mathbf{p}$ with
$\mathbf{p}_{update}$ and check the power of the updated
perturbation vector. If it is below the desired perturbation power
we keep it, otherwise, we normalize it to meet the power
constraint. Using this iterative procedure, we are able to craft a
universal perturbation $\mathbf{p}$, which is independent of the
input of the autoencoder.

\begin{figure}[]
	\centering
	\includegraphics[width= 1 \columnwidth, trim={0.3cm 0.2cm 0.9cm 1.15cm},clip]{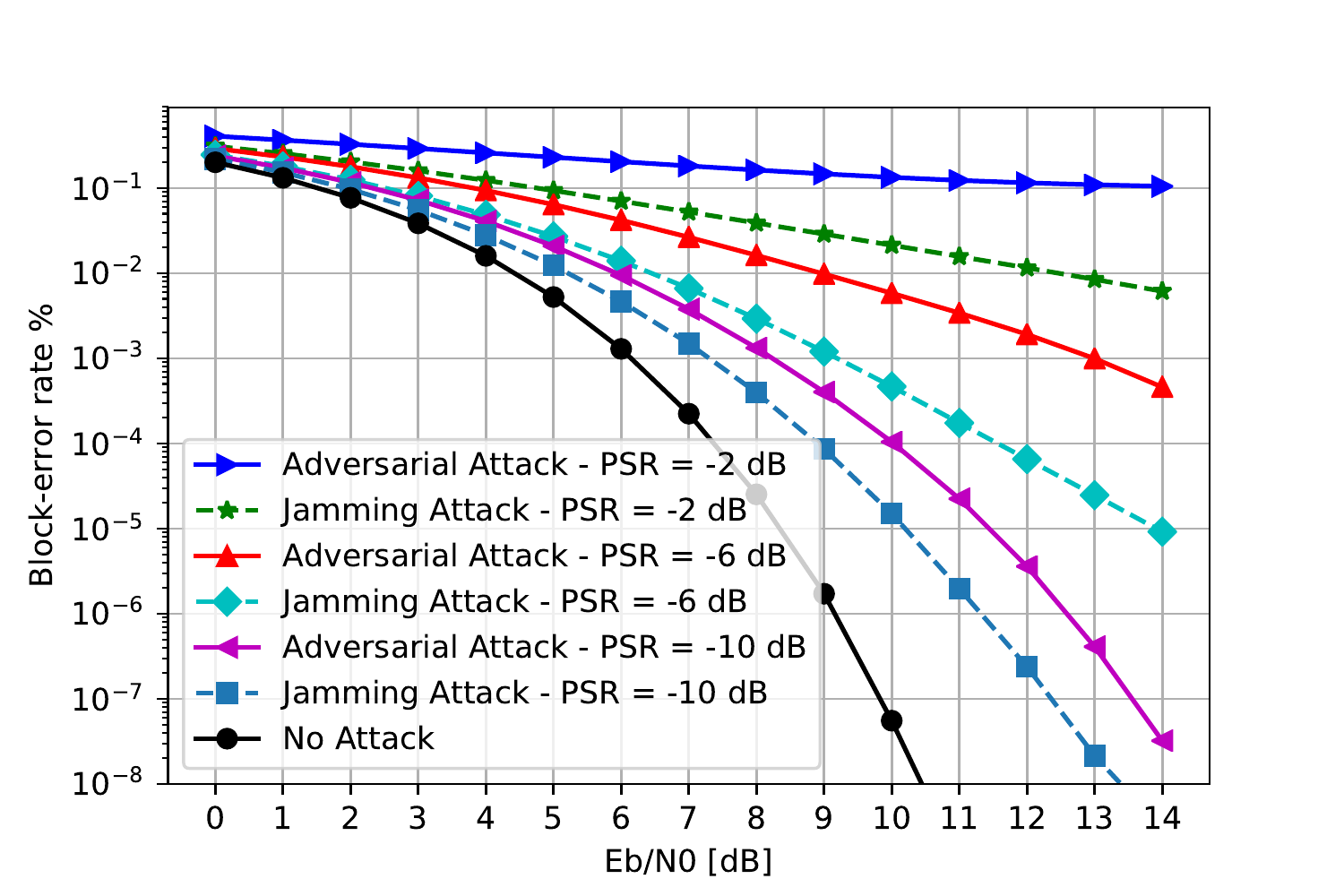}
	\caption{BLER versus $E_{b}/N_{0}$ under Adversarial and Jamming attacks against the MLP autoencoder in Table I.}
	\label{fig2}
\end{figure}

Figure \ref{fig2} presents the block-error-rate (BLER) performance of the MLP autoencoder of Table \ref{table1}, under adversarial attack. The adversarial attack is designed using Alg. \ref{alg1}. To compare the power of the adversarial perturbation at the receiver with the received signal power, we define a metric called
perturbation-to-signal ratio (PSR), which is equal to the ratio of the received perturbation power to the received signal power. For the sake of comparison, we also consider a classical jamming attack, where a jammer creates Gaussian noise with the same PSR as the adversarial attack. Gaussian jamming is the most effective way to reduce the mutual information for a message encoded with a capacity-achieving Gaussian codebook (which is optimum for the AWGN channel), under a given jamming power constraint. Based on Fig. \ref{fig2} it is clear that even for small PSR values, the adversarial attack can degrade the BLER by
orders of magnitude. Also, the adversarial attack has a more destructive impact compared to the jamming attack.

\section{Crafting Shift-Invariant Black-Box Attacks}
In Section III, we considered two restrictive assumptions in order to craft adversarial perturbations. First, we assumed that the adversarial attacker has  perfect knowledge of the autoencoder system, including the number of layers, weight and bias parameters of the decoder block. Second, the attacker is synchronous with the transmitter. More precisely, given that $\mathbf{p}$ and $\mathbf{x}$ are $2n$ dimensional real vectors ($n$-dimensional complex vectors), then $\mathbf{p} + \mathbf{x} + \mathbf{n}$ requires each element of $\mathbf{p}$ to be added by its corresponding element of $\mathbf{x}$. We address these problems by applying a heuristic approach and the transferability of adversarial attacks.

Consider the CNN autoencoder of Table \ref{table1}, and assume an attacker is interested to create an adversarial attack for it, without having any knowledge about its structure or parameters. The transferability of adversarial attacks says that attacks designed for a specific model are also effective for other models, with high probability \cite{Papernot2016Transferability}. Therefore, the attacker can use its own substitute autoencoder communication system and then design a white-box attack for it, as it has the perfect knowledge of this substitute autoencoder communication system. Then use the designed perturbation to attack the original unknown model. This attack is called a black-box adversarial attack \cite{Papernot2016Transferability}. 

\begin{algorithm}[t]
	
	\caption{ Crafting Shift-Invariant Perturbations \label{alg2}}
	\begin{algorithmic}[1]
		\State Using the substitute network, generate $I$ adversarial perturbations using Alg. \ref{alg1}.
		
		\State Calculate the BLER of a randomly shifted version of each of the $I$ perturbations on the substitute network.
		
		\State Select the first $t$ perturbations associated with the $t$ least BLERs. Denote them as $\{ \mathbf{p}_{1}, \ldots, \mathbf{p}_{t} \}$.
		
		\State Set $\mathbf{P}_{norm} = \left[ \dfrac{\mathbf{p}_{1}}{\Vert \mathbf{p}_{1} \Vert_{2}}, \dfrac{\mathbf{p}_{2}}{\Vert \mathbf{p}_{2} \Vert_{2}}, \ldots, \dfrac{\mathbf{p}_{t}}{\Vert \mathbf{p}_{t} \Vert_{2}} \right]^{T}$.
		
		\State Calculate the SVD of $\mathbf{P}_{norm}$ as $\mathbf{P}_{norm} = \mathbf{U} \mathbf{\Sigma} \mathbf{V}^{T}$.
		
		\State Select the first column of $\mathbf{V}$ as the candidate shift-invariant perturbation, i.e., $\mathbf{p}_{si} = \mathbf{V} \; \mathbf{e}_{1}$.
	\end{algorithmic}
\end{algorithm}

Using this approach, we consider the MLP autoencoder in Table~\ref{table1} as the substitute model and create an adversarial perturbation for it. Then, we use the so-created perturbation to attack the CNN autoencoder in Table~\ref{table1}, which is unknown to the attacker. Note that this approach is general and we can use it to attack other autoencoders with different structures and parameters.

In order to remove the synchronicity requirement between the attacker and the transmitter, we present a heuristic algorithm, Alg. \ref{alg2}, to create attacks which are robust against random time shifts. This is a challenging task due to the considered setup: $\!n\!=\!\!7$ defines a low-dimensional input, whereas generally, finding effective adversarial attacks is easier the larger the dimensions of the input and the model are \cite{goodfellow2014explaining}.

The main ideas is as follows. Using the substitute network, first we generate a pool of adversarial perturbations, e.g., $I$ adversarial perturbations, using Alg. 1. Then, for each of these $I$ perturbations, we calculate the BLER of a randomly shifted version of them. Next, we rank them based on the severity of their attacks. Using this approach, the top $t$ perturbations show a robustness against random shifts. Note that each of these $t$ perturbations represents a direction in the feature space that even a randomly shifted version of it causes significant misclassification. Therefore, we can use a singular value decomposition (SVD) of these $t$ vectors to find their main principal direction which hopefully would show a better shift-invariant property. More details are given in Alg. \ref{alg2}. Using the proposed algorithm, we are able to craft adversarial attacks which are robust against random shifts, and therefore a synchronous attack is not required.

\begin{figure}[]
	\centering
	\includegraphics[width= 1 \columnwidth, trim={0.3cm 0cm 0.9cm 0.7cm},clip]{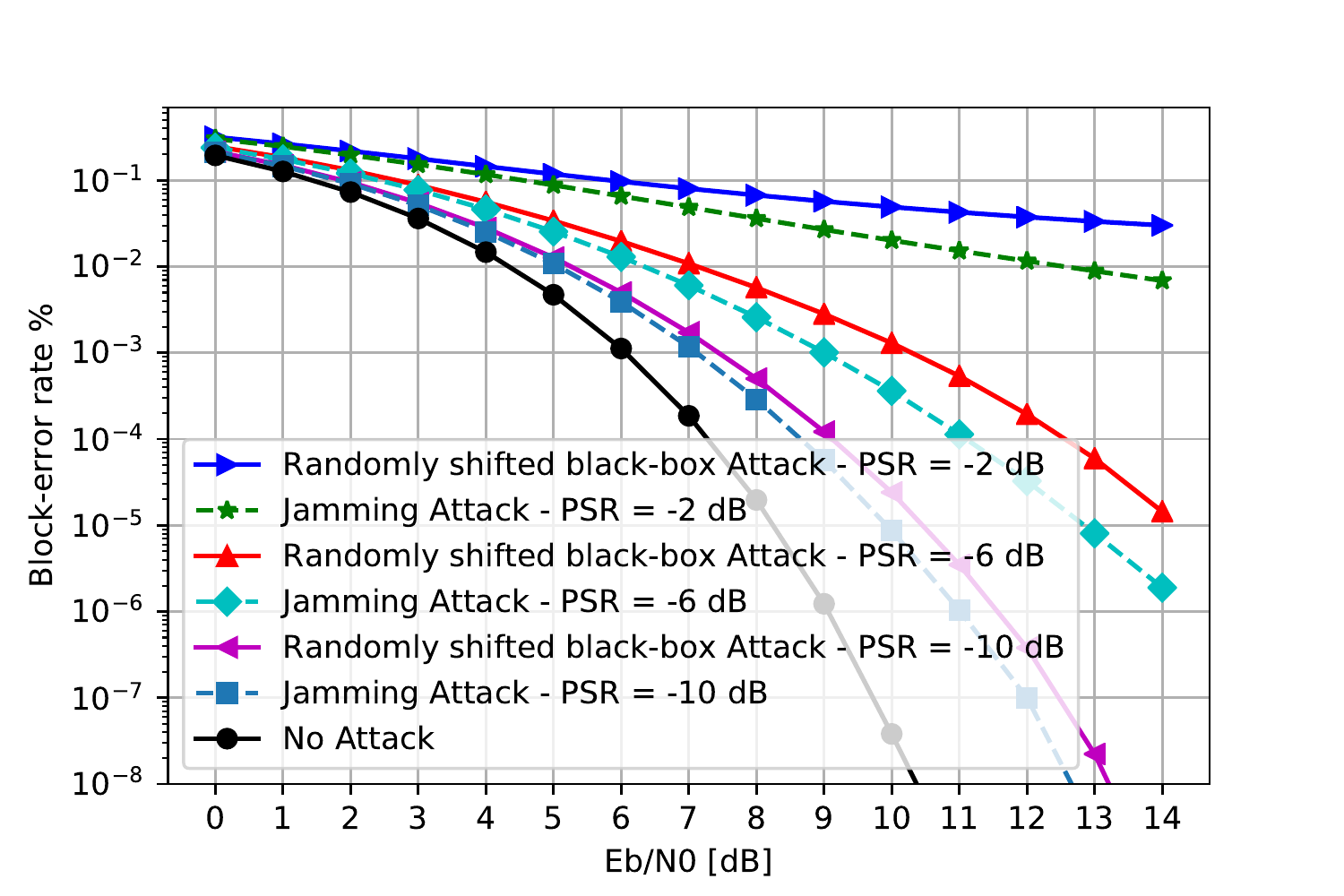}
	\caption{BLER versus $E_{b}/N_{0}$ for randomly shifted black-box adversarial attacks against the CNN autoencoder in Table I.}
	\label{fig3}
\end{figure}

Using a black-box attack and the heuristic method proposed in Alg. \ref{alg2}, we attack the CNN autoencoder of Table~\ref{table1}, while we use the parameters of the MLP autoencoder of Table~\ref{table1}. The result is shown in Fig. \ref{fig3}. Fig. \ref{fig3} reveals two important properties of non-synchronous black-box \emph{physical} adversarial attacks against end-to-end autoencoder communication systems. First,  they are significantly effective. Second, they can provide a more destructive effect compared to jamming attacks.\footnote{In numerical experiments not included here due to space limitations, we simulated 10 different models (with different weights and biases) and applied the same attack to them. We observed the same trend as before, suggesting that our conclusions are general and not specific to a particular model \cite{code_mey_ae}.} 

\section{Robustness of Autoencoders vs Classical Approaches}
\begin{figure}[]
	\centering
	\includegraphics[width= 1 \columnwidth, trim={0.45cm 0.5cm 1cm 1cm},clip]{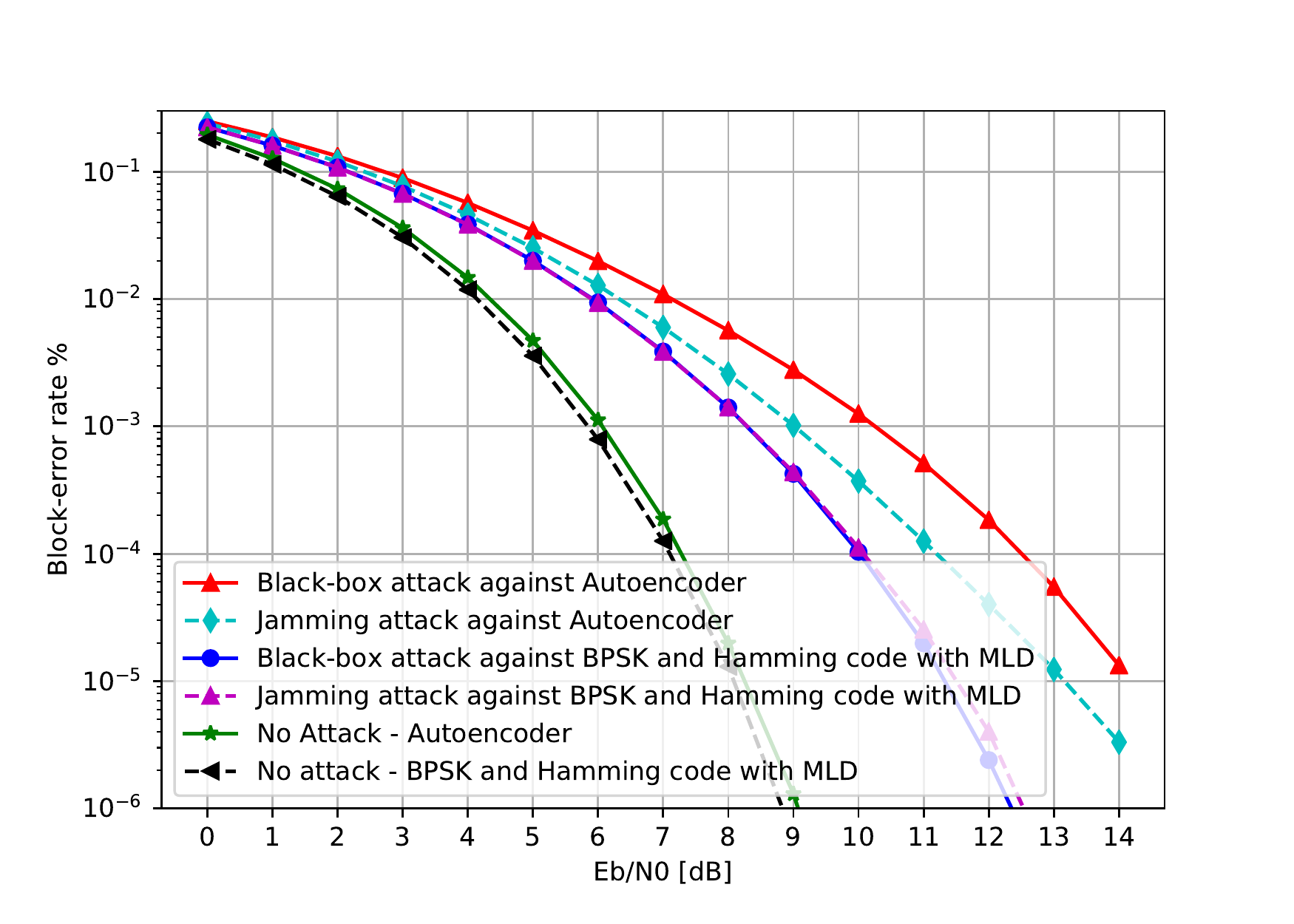}
	\caption{Comparison of the robustness of
		the autoencoder (CNN, see Table I) and classical
		transmission (BPSK modulation with Hamming coding). The PSR is set to $-6$ dB.}
	\label{fig4}
\end{figure}

In this section, we compare the robustness of the end-to-end
autoencoders with a classical modulation and coding
scheme. In \cite{TimDL4Phy}, it has been shown that
the end-to-end autoencoder provides roughly the same BLER as binary
phase-shift keying (BPSK) modulation combined with a Hamming (7,4)
code and maximum-likelihood decoding (MLD). This also can be
verified from Fig.~\ref{fig4}. We use BPSK modulation and Hamming
(7,4) coding with MLD as a benchmark and compare the robustness of
the CNN autoencoder in Table \ref{table1} with it. For the attack,
we use both the jamming attack and the shift invariant black-box
attack of Section III. The results are presented in Fig. \ref{fig4},
where the PSR for both attacks is equal to $-6$ dB.

From Fig. \ref{fig4}, we make the following
observations. First, the randomly shifted black-box attack causes a
bigger BLER than a jamming attack for the considered autoencoder
system. Second, the BPSK modulation with Hamming coding and MLD
results in roughly the same BLER for both adversarial and jamming
attacks. Third, the classical BPSK modulation with Hamming coding and
MLD provides a smaller BLER than the considered CNN autoencoder
under a jamming attack. Therefore, Fig. \ref{fig4} suggests that the
performance of the  classical approach is more robust, compared to
autoencoders, against both adversarial attacks and jamming attacks.

\section{Conclusions, Future Works and Further Discussions}
We showed end-to-end autoencoder systems are significantly vulnerable to small perturbations, which can decrease their BLER by orders of magnitude. Due
to the broadcast nature of the wireless channel, creating such perturbations is easy for an adversarial attacker. We also presented algorithms to craft such \emph{physical} adversarial attacks and verified them by simulations. These findings suggest that defense mechanisms against adversarial attacks and further research on the security and robustness of deep-learning based wireless systems is a necessity.

One possible defense mechanism is to train the
autoencoder with adversarial perturbations (which is a technique to
increase robustness and is known as adversarial training
\cite{goodfellow2014explaining}). However, adversarial training has
its own drawbacks, e.g., the attacker may use a different attack
than the one used for training the network. Also, the attacker can
design adversarial perturbations for an autoencoder that already has
been trained with adversarial training, and craft new adversarial
perturbations. Also, adversarial training can reduce the performance
of the autoencoder on clean inputs.

We also compared the robustness of a CNN-based end-to-end autoencoder system with BPSK modulation and Hamming coding and MLD. We showed that the classical approach provides more robust performance against both adversarial and jamming attacks, compared to the end-to-end autoencoder.

Our results were obtained for physical adversarial attacks in AWGN channels, considering AWGN jamming. Further studies are required to investigate other channel models (e.g., Rayleigh fading), a wider set of hyperparameters and information rates (e.g., we only studied $n=7$ and $k=4$), and comparisons against other, more advanced, jamming strategies (see, e.g. \cite{LiuAntiJamming}). For example, it is an open problem to determine more exactly, quantitatively, how the gap between the jamming attack and the adversarial attack behaves as function of the information rate. Also, the extension to more advanced channel models, e.g., Rayleigh fading, is an interesting direction for future work. In that case, the attacker must have an estimate of the channel between himself and the receiver. The quality of this estimate may fundamentally affect the degree of success of the attacker. We hope that our initial results will stimulate future research on this topic.

\bibliographystyle{IEEEtran}

\bibliography{IEEEabrv,DLandWCAE}

\end{document}